\documentclass[amsmath,amssymb,superscriptaddress,nobalancelastpage,prb,twocolumn]{revtex4-1}

\usepackage{hyphenat}
\usepackage{graphicx,xcolor}
\usepackage{graphicx}
\usepackage{varioref}
\usepackage{xr-hyper}
\usepackage{xcolor}
\usepackage{nicefrac}
\usepackage{xfrac}
\usepackage{hyperref}
\hypersetup{colorlinks,linkcolor=blue,urlcolor=blue,citecolor=blue}
\usepackage{ulem}
\usepackage{siunitx}
\usepackage{graphicx}
\usepackage{dcolumn}
\usepackage{bm}
\usepackage{braket}
\usepackage{wasysym}
\usepackage{textcomp}
\usepackage{mhchem}

\usepackage{braket,diagbox,varioref,xr-hyper,nicefrac,xfrac,dcolumn,textcomp}
\newcommand{\cref}[1]{Chap.~\ref{#1}}

\makeatletter
\newlength{\Laped}
\newcommand{\lscox}[1]{\setlength{\Laped}{2pt-#1pt}\ce{La_{\strip@pt\Laped}Sr_{#1}CuO4}}

\makeatother
\newcommand{\ybco}{YBa$_2$Cu$_3$O$_{6+y}$}
\newcommand{\pbco}{PrBa$_2$Cu$_3$O$_{7}$}
\newcommand{\ypbco}{Y$_{0.7}$Pr$_{0.3}$Ba$_2$Cu$_3$O$_{6.67}$}
\newcommand{\ypbcoseven}{Y$_{0.7}$Pr$_{0.3}$Ba$_2$Cu$_3$O$_{7}$}
\newcommand{\gypbco}{Y$_{1-x}$Pr$_{x}$Ba$_2$Cu$_3$O$_{6+y}$}

\begin{document}



\title{Differentiation of Site-Specific Symmetry Breaking Orders in Y$_{1-x}$Pr$_x$Ba$_2$Cu$_3$O$_{6+y}$}

\author{L. Martinelli}
\affiliation{Physik-Institut, Universit\"{a}t Z\"{u}rich, Winterthurerstrasse 190, CH-8057 Z\"{u}rich, Switzerland}

\author{S. Rüdiger}
\affiliation{Physik-Institut, Universit\"{a}t Z\"{u}rich, Winterthurerstrasse 190, CH-8057 Z\"{u}rich, Switzerland}

\author{I. Biało}
\affiliation{Physik-Institut, Universit\"{a}t Z\"{u}rich, Winterthurerstrasse 190, CH-8057 Z\"{u}rich, Switzerland}

\author{J.~Oppliger}%
\affiliation{Physik-Institut, Universit\"{a}t Z\"{u}rich, Winterthurerstrasse 
190, CH-8057 Z\"{u}rich, Switzerland}%

\author{F.~Igoa Saldaña}
\affiliation{Deutsches Elektronen-Synchrotron DESY, Notkestra{\ss}e 85, 22607 Hamburg, Germany.}

\author{M.~v.~Zimmermann}
\affiliation{Deutsches Elektronen-Synchrotron DESY, Notkestra{\ss}e 85, 22607 Hamburg, Germany.}

\author{E.~Weschke}
\affiliation{Helmholtz-Zentrum Berlin für Materialien und Energie, Albert-Einstein-Strasse 15, D-12489 Berlin, Germany.}


\author{R. Arpaia}
\affiliation{
Department of Molecular Sciences and Nanosystems, Ca’ Foscari University of Venice, I-30172, Venice, Italy
}
\affiliation{
Quantum Device Physics Laboratory, Department of Microtechnology and Nanoscience, Chalmers University of Technology, SE-41296 Göteborg, Sweden
}

\author{J.~Chang}
\affiliation{Physik-Institut, Universit\"{a}t Z\"{u}rich, Winterthurerstrasse 190, CH-8057 Z\"{u}rich, Switzerland}

\date{\today}

\begin{abstract}
\end{abstract}


\maketitle

\textbf{Solid matter is classified through symmetry of ordering phenomena. Experimentally, this approach is straightforward, except when distinct orderings occur with identical or almost identical symmetry breaking. Here we show that the cuprate system Y$_{1-x}$Pr$_x$Ba$_2$Cu$_3$O$_{6+y}$ hosts two distinct orderings with almost identical translational symmetry breaking. Only when applying site-sensitive resonant elastic x-ray scattering (REXS), charge ordering can be conclusively differentiated from a super-lattice structure. These two orderings occur with almost the same in-plane symmetry but manifest at different atomic sites and display different temperature dependence. Differentiating these orders provides an important clue to the anomalous behavior of PrBa$_2$Cu$_3$O$_7$ within the 123-series of high-temperature superconductors.
We conclude that the symmetry breaking at the Pr-site is unfavorable for superconducting pairing. }


The high-temperature superconducting compound family REBa$_2$Cu$_3$O$_{6+y}$ (REBCO) is used to enhance the performance of critical technologies such as windmill rotors~\cite{bergen_design_2019}, fusion tokamaks~\cite{li_rebco_2025}, and magnets~\cite{dong_construction_2025} including MRI~\cite{mukoyama_superconducting_2018} devices. 
Generally, these compounds display 
similar superconducting onset temperatures and magnetic ordering at very low temperatures -- irrespectively of the rare-earth (RE) element.
However, PrBa$_2$Cu$_3$O$_{6+y}$ (PBCO) represents a puzzling anomaly.
Although maintaining the same average crystal structure, PBCO is not superconducting or even metallic. Moreover, the Pr ions order antiferromagnetically at temperatures one order of magnitude larger ($\sim12-17$ K) than other RE ions~\cite{kiss_theoretical_2010}. 
Therefore, PBCO has been the subject of theoretical scrutiny. 
The most accepted theoretical model that explains the suppression of superconductivity was developed by Fehrenbacher and Rice (FR)~\cite{fehrenbacher_unusual_1993}. It revolves around the hybridization of the Pr-$4f$ and O-$2p$ orbitals, which -- in turn -- impedes efficient doping of the CuO$_2$ planes. 
The FR model considers a local hole state formed by the coordination of O-$2p_\pi$ orbitals oriented towards the Pr site with $4f_{z(x^2-y^2)}$ symmetry. The result is a $4f^1 + 4f^2 \underline{L}$ Pr configuration ($\underline{L}$ stands for a hole in the oxygen ligand bands), so that Pr has a mixed $3^+$/$4^+$ valence.
This model was able to explain some of the experimental results, including inelastic neutron scattering, X-ray absorption, and magnetic susceptibility~\cite{merz_x-ray_1997,soderholm_crystal-field_1991}. 

An important question relates to electronic ordering in PBCO. Charge ordering~\cite{ghiringhelli_long-range_2012,chang_direct_2012,achkar_distinct_2012,blackburn_x-ray_2013,hucker_competing_2014,choi_spatially_2020,vinograd_using_2024,blanco-canosa_momentum-dependent_2013,blanco-canosa_resonant_2014,kim_charge_2021,wu_magnetic-field-induced_2011,forgan_microscopic_2015,vinograd_locally_2021,wu_emergence_2013,wu_incipient_2015,leboeuf_thermodynamic_2013, wahlberg_restored_2021} and associated electronic reconstruction~\cite{doiron-leyraud_quantum_2007,chang_nernst_2011,rourke_fermi-surface_2010,ramshaw_quasiparticle_2015} have been studied in great detail in YBa$_2$Cu$_3$O$_{6+y}$ (YBCO). Generally, a two-dimensional, bi-directional charge-density-wave (CDW) order is found in a doping range centered around the so-called 1/8-anomaly~\cite{hucker_stripe_2011,julien_magnetic_2003,chang_tuning_2008}. Upon application of magnetic field~\cite{gerber_three-dimensional_2015,chang_magnetic_2016} or uniaxial pressure~\cite{kim_uniaxial_2018}, a three-dimensional uni-directional charge order was reported. The two types of CDW structures seem to be linked as they occur with exactly the same in-plane ordering vector. Furthermore, both seem to compete with superconductivity. More recently, evidence of different CDW patterns in compounds with partial $Y\rightarrow Pr$ has been reported~\cite{ruiz_stabilization_2022,kang_discovery_2023}.  However, studies focusing on PBCO have received little attention.
 
\begin{figure*}
    \center{\includegraphics[width=0.999\textwidth]{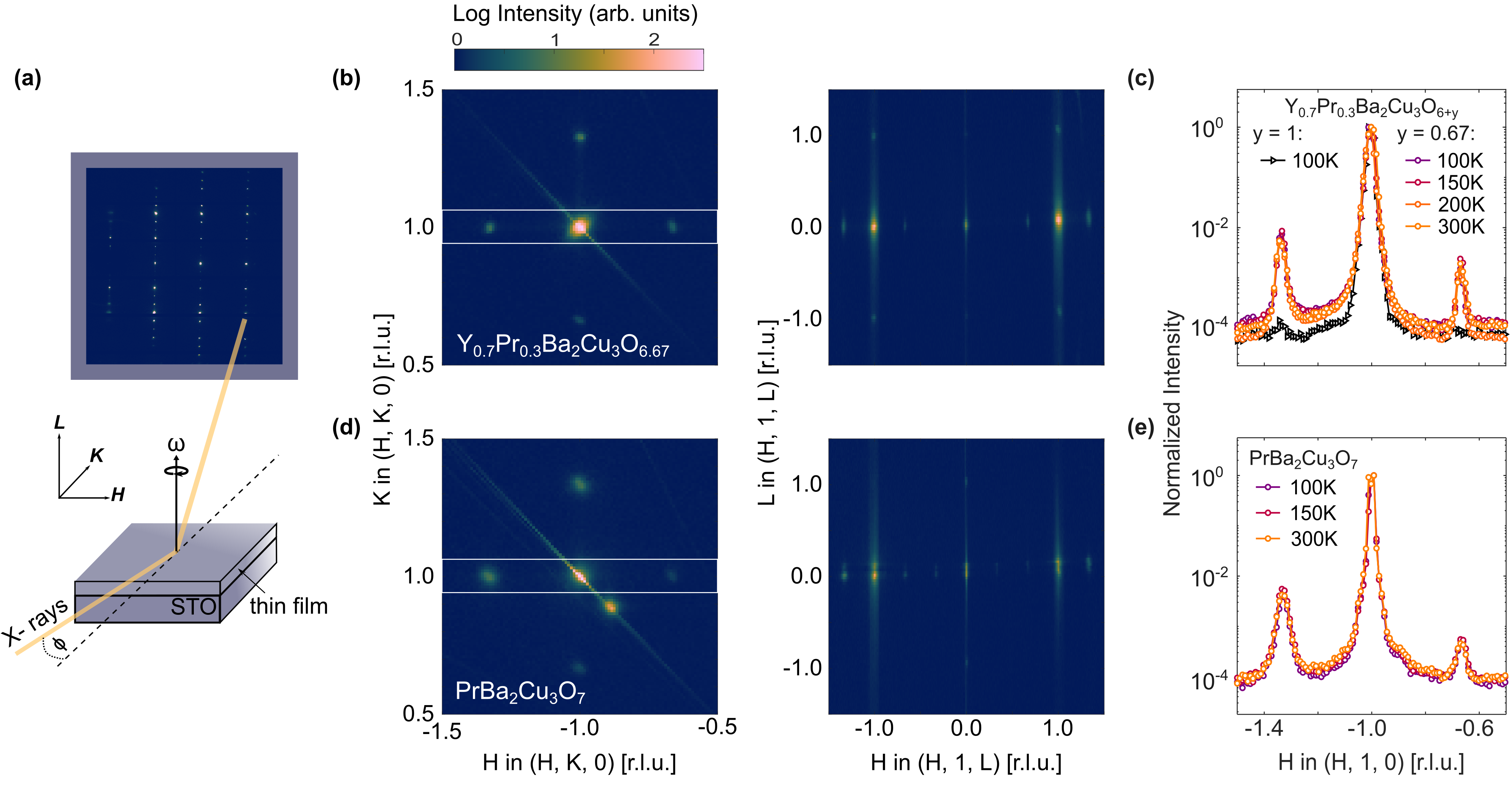}} 
    \caption{\textbf{Grazing-incidence x-ray diffraction on \pbco\ and \ypbco\ thin films.} (a) Schematic illustration of the scattering technique. The x-ray incidence angle $\phi$ has been kept fixed. The films are rotated continuously by the angle $\omega$ around the axis normal to the film surface, while detector images are recorded at fixed framerate. The actual value of $\omega$ is measured for each frame. (b,d) Scattering volumes -- represented by the $(H,K,0)$ and $(H,1,L)$ scattering planes -- recorded on \pbco\ and \ypbco\ thin films. The scattering intensity is displayed with a logarithmic false color scale. (c,e) One-dimensional intensity scans along $(H,1,0)$ derived from the data shown in (b,d) and for temperatures as indicated, normalized to the intensity of the $(-1,1,0)$ Bragg reflection. In (c), a scan recorded on \ypbcoseven\ is included. }
    \label{fig:figure1}
\end{figure*}


Here, we present a comprehensive study of \pbco\ and \gypbco\ ($x=0.3$, $y =0.67$ and 1) films. Lattice and charge modulations are revealed by a combination of high-energy grazing incidence x-ray diffraction (GI-XRD) and resonant elastic x-ray scattering (REXS). Across the $(x,y)$ phase diagram, we identify two ordering vectors: $Q_1=(\delta_1,0,0)$ and $Q_2=(\delta_2,0,1/2)$ with $\delta_1\approx \delta_2\approx 1/3 $.  Although these two orders display	 very similar in-plane periodicity, we demonstrate that they stem from different mechanisms. This conclusion is based on the following observations. (1) They resonate at different atomic sites. $Q_1$ resonates at the Pr and out-of-plane copper sites, whereas $Q_2$ resonates at the in-plane copper site. (2) The two order parameters display different temperature dependences. The $Q_2$-order vanishes with temperature whereas the $Q_1$ one is essentially temperature independent.
(3) The two orders display different correlation lengths. $Q_1$ is a long-range order comparable to the average atomic lattice, whereas $Q_2$ is short-range correlated. 
(4) The in-plane periodicity shows a minute discrepancy, in that we find $\delta_2<1/3<\delta_1$. We interpret the $Q_2$ order in terms of a charge-density-wave identical to what has been reported in YBCO. The $Q_1$ order -- most pronounced in PBCO and absent in YBCO -- is interpreted as a Pr-related super-lattice structure. This structure is likely the key to understanding the unusual electronic properties of PBCO.

\begin{figure*}
    \center{\includegraphics[width=0.999\textwidth]{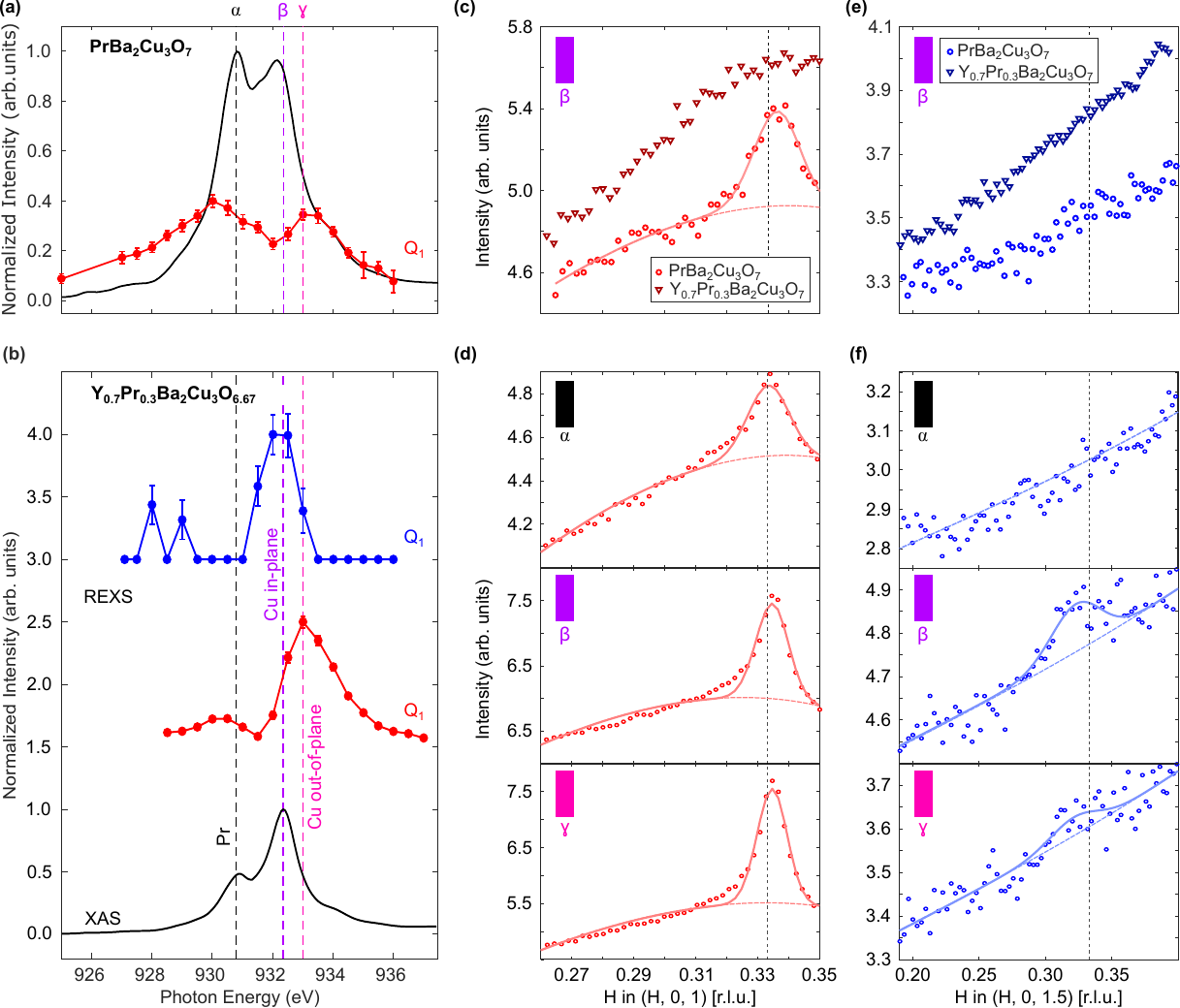}} 
    \caption{\textbf{X-ray absorption spectroscopy (XAS) and resonant elastic x-ray scattering (REXS).} (a,b) X-ray absorption spectroscopy (XAS) -- black curves -- across the Pr and Cu-L resonances (indicated by vertical dashed lines) on \pbco\ and \ypbco.   The REXS signal at $(1/3, 0, 1)$ and $(1/3, 0, 1.5)$ shown with respectively red and blue points.  (c-f) Momentum dependence of the REXS signal through $(1/3,0,1)$ and $(1/3, 0, 1.5)$ at respectively the in- and out-of-plane Cu-$L$ resonances ($T=10$~K). (c) and  (e) include scans recorded on \ypbcoseven. Solid lines represents Gaussian fits to the data with dashed lines indicating the background of these fits. } 
    \label{fig:figure2}
\end{figure*}

\begin{figure*}
    \center{\includegraphics[width=0.999\textwidth]{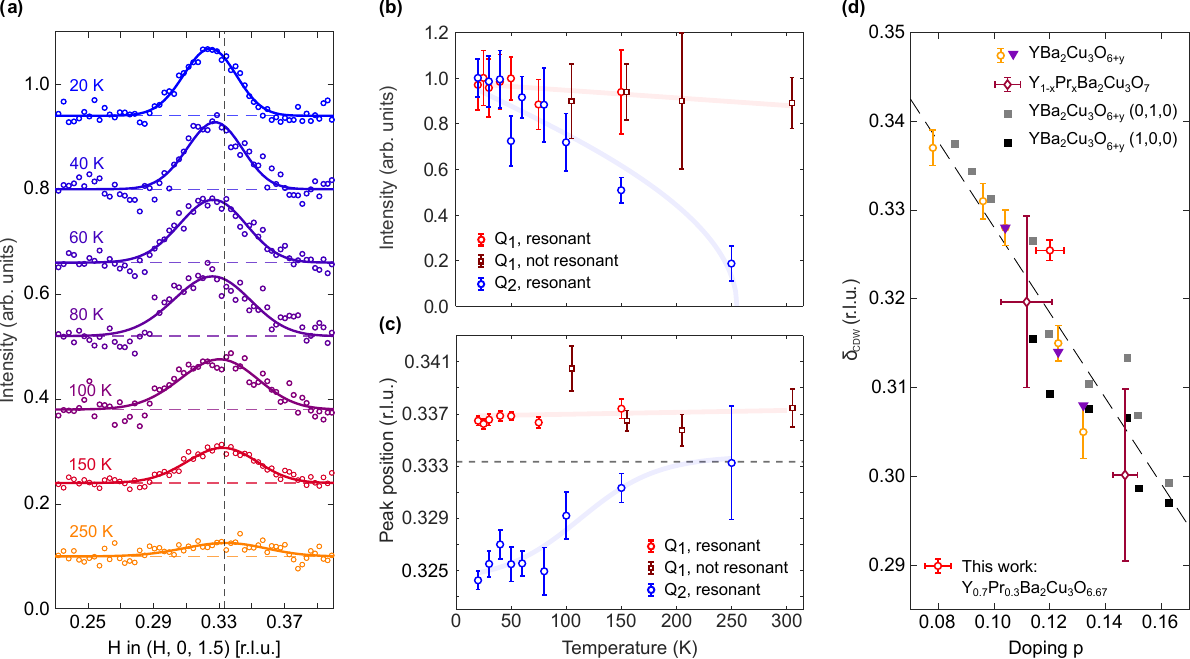}} 
    \caption{\textbf{Temperature dependence of the charge order in \ypbco.} (a) Background subtracted intensity versus momentum across the charge ordering vector in \ypbco, for temperatures as indicated. Solid and dashed lines are respectively Gaussian fits and the linear background. (b) Intensity (amplitude) of the charge-density-wave order reflection (blue) compared with the $L=0$ modulation. (c) Incommensurabilities of the two modulations found in \ypbco\ as a function of temperature. Error bars in (b,c) indicate standard deviations of the Gaussian fits. (d) 
    Charge-density-wave incommensurability (Ref.~\onlinecite{blanco-canosa_resonant_2014,kang_discovery_2023,blackburn_x-ray_2013,hucker_competing_2014}) in YBCO and \ypbco. Dashed lines are guides to the eye. }
    \label{fig:figure3}
\end{figure*}

\section{Results}{\label{sec_res}}
\textit{Non-resonant x-ray scattering:}  The combination of the high-energy x-ray diffraction with grazing-incidence geometry provides a reciprocal space overview with high sensitivity to the sample surface layers~\cite{gustafson_high-energy_2014}.  In this work,
we take a further step and combine high-energy grazing-incidence diffraction with low
temperatures to investigate structural orderings in
thin films of \pbco, \ypbcoseven\, and \ypbco. In Fig.~\ref{fig:figure1}, $(H,K,0)$ and $(H,1,L)$ scattering planes are obtained by slicing the three-dimensional reciprocal scattering volume.
The most intense reflections are associated with fundamental Bragg peaks at $Q_B=(H,K,L)$ with integer Miller indices. Furthermore, quasi-commensurate weaker reflections are found at $Q_1=Q_B+\tau_1$ with $\tau_1\approx(\pm\delta_1,0,0)$ and  $(0,\pm \delta_1,0)$ where $\delta_1\gtrapprox 1/3$. These quasi-commensurate reflections are two to four orders of magnitude weaker than the fundamental Bragg reflections in \pbco\ and \ypbco\ -- see Fig.~\ref{fig:figure1}c,e. In \ypbcoseven, these reflections are yet another one to two orders weaker and are only observed in a few Brillouin zones. In all cases, the $Q_B$ and $Q_1$ reflections are essentially temperature independent in the explored temperature range from 100 to 300~K (see Fig.~\ref{fig:figure1}c).
The $Q_1$ peaks display a full-width-at-half-maximum (FWHM) similar to the Bragg peaks in both \ypbco~and \pbco, indicating that the width is dominated by the experimental resolution.

\textit{Site-sensitive measurements:} Resonant scattering experiments were carried out on both the Cu-$L$ and Pr-$M$ resonances. As already revealed by the non-resonant grazing incident diffraction, the crystal structure includes a $\tau_1$ modulation.  
With resonant scattering, the quasi-commensurate reflection is probed through $Q_1~=~(\delta_1,0,1)$.
In this zone, the reflection is too weak to be observed in \ypbcoseven, but intense reflections are found in \pbco\ and \ypbco. The photon-energy dependence of this commensurate peak displays maxima at the Pr and out-of-plane Cu resonances 
(see Fig.~\ref{fig:figure2}a,b).  The peak positions -- determined through Gaussian fits 
-- are, respectively $\delta_1=0.3365(4)$ for \ypbco\ and $\delta_1=0.337$ for \pbco. Thus, for both compounds $\delta_1>1/3$. Just as with non-resonant scattering, these reflections are essentially temperature independent.  

\textit{Charge density wave order:} Only in \ypbco, a much weaker reflection is found at $Q_{2}=(\delta_{2},0,1/2)$ with $\delta_{2}=0.325(1)<1/3$. This reflection is most intense at the in-plane Cu-$L$ resonance (Fig.~\ref{fig:figure2}b). Furthermore, it displays a significant temperature dependence, as shown in Fig.~\ref{fig:figure3}a, and essentially disappears at 250 K.
 The peak amplitude -- quantified through Gaussian fits -- decreases with increasing temperature (Fig.~\ref{fig:figure3}b). At high temperature (with weakened amplitude), the peak position slightly shifts to larger momentum (Fig.~\ref{fig:figure3}c) -- as previously observed in La- and Bi-based cuprates~\cite{wang_high-temperature_2020,arpaia_signature_2023}. 
 Importantly, the $Q_2$ reflection shows little dependence on the out-of-plane momentum $L$ (Fig.~S4 in the Supplemental Material), indicating a predominant two-dimensional character.
The FWHM shows a typical trend for charge order, decreasing with temperature and plateauing below the onset of superconductivity, as shown in Fig.~S3 in the Supplemental Material. The maximum correlation length, calculated as $\frac{a}{\pi\cdot \text{FWHM}}$, is around $30$~\AA. This value is approximately two times smaller than the correlation length in \ybco \cite{blanco-canosa_resonant_2014}.
 We also stress that, compared to YBCO, the charge density wave peak has a significantly higher onset temperature. Yet, the low-temperature commensuration is consistent with that of charge-density-wave order found in YBCO for the same oxygen concentration (Fig.~\ref{fig:figure3}d). 


\begin{figure}
    \center{\includegraphics[width=0.4999\textwidth]{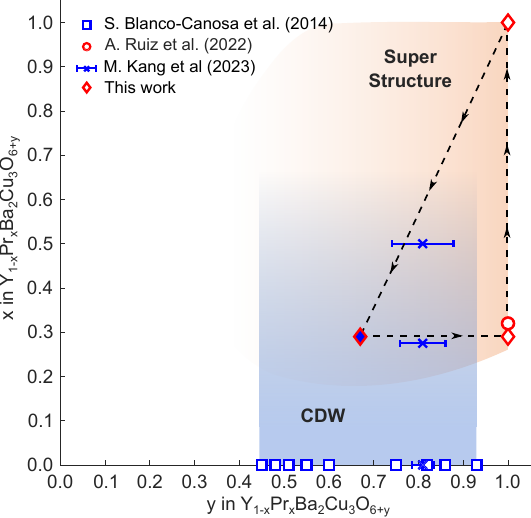}}
    \caption{\textbf{Structural phase diagram of \gypbco.} Charge order (blue) and super lattice structure (orange) depend on oxygen content and Pr substitution. Red and blue symbols indicate respectively the observation of super-structure and charge density wave order. Color shaded areas are schematically showing the extend of the two phases. }
    \label{fig:figure4}
\end{figure}
\section{Discussion}
In \ybco, charge orders with integer and half-integer out-of-plane modulation have been observed with exactly the same in-plane ordering vector~\cite{gerber_three-dimensional_2015,chang_magnetic_2016,jang_ideal_2016}.
A schematic phase diagram of the different peaks observed in this study and recent literature (Refs.~\cite{kang_discovery_2023, ruiz_stabilization_2022}) as a function of oxygen and Pr doping is reported in Fig.~\ref{fig:figure4}.
Due to the identical in-plane periodicity, these orders have been interpreted in terms of different stacking patterns of the same in-plane charge-density wave order. 
The observation, in \gypbco, of integer and half-integer out-of-plane modulations with almost the same in-plane ordering vectors are reminiscent of what has been found in YBCO~\cite{ruiz_stabilization_2022}.
However, there are at least three important differences. (i) Although close, the in-plane ordering vectors (in \ypbco) are not identical. (ii) The two orders are resonating at different atomic sites. (iii) The integer out-of-plane ordering displays no temperature dependence. For these reasons, we will discuss the two orderings separately.

The chemical inclusion of partial Pr-substitution may influence the distribution of doped holes. 
In YBCO, an oxygen content of 6.67 corresponds to 0.12 holes per in-plane copper site~\cite{liang_growth_2012,liang_evaluation_2006}. 
The relatively low superconducting transition temperature ($T_c=30$~K) in our \ypbco\ thin film 
may indicate a reduced hole doping $p_{\textrm{ZR}}$ within the CuO$_2$ planes. The remaining holes $p_{\textrm{FR}}$ would be residing within the Pr $4f$-FR state (so that $p=p_{\textrm{ZR}}+p_{\textrm{FR}}$)~\cite{merz_x-ray_1997}.
The charge-density-wave incommensurability suggests a hole doping of $p_\textrm{ZR}\approx0.10$ within the CuO$_2$ planes. As such, our results suggests that only a small fraction of holes ($p_\textrm{FR}=0.02$) goes into the Pr $4f$-FR state, and that these holes do not participate in the charge order modulation.

The weak reflection occurring at $Q_{2}=(\delta_2,0,1/2)$ -- in  \ypbco -- is here interpreted as evidence of charge-density-wave order.
Phenomenologically, this order resembles the two-dimensional charge-density-wave order observed in YBCO. It has the same ordering vector, resonates at the in-plane copper site, and shows a similar decrease with temperature. The larger onset temperature of charge-density-wave order combined with a lower superconducting transition temperature could be interpreted in terms of phase competition. 
Below the superconducting critical temperature, a saturation of the charge-density wave intensity is observed, similar as in other cuprates~\cite{comin2016Resonant}.



Next, we turn to the $Q_1$-order manifested by quasi-commensurate peaks at $Q_1=(\delta_1,0,0)$.
For \pbco, $\delta_1>1/3$ whereas charge-density-wave ordering in optimally doped YBCO has an incommensurability $\delta_{2}<1/3$~\cite{hucker_competing_2014,blanco-canosa_resonant_2014}. The diffraction intensity of the $Q_1$ peaks and the absence of temperature dependence furthermore are at odds with the expectations for an electronic order. In what follows, we discuss this ordering in terms of a super-lattice structure. Given that this ordering resonates at the out-of-plane Cu and Pr sites, this suggests that it is not linked to the 
in-plane Cu orbitals.
We observe this super-lattice structure in both \pbco, \ypbcoseven\ and \ypbco. In \ypbcoseven\ and \ypbco, the 1/3 partial substitution provides a direct motive for such a superstructure. A Pr-substitution every third unit cell would generate peaks at $Q_1$. 
Given that the $Q_1$ order manifest strongest in \pbco\ and weakest in \ypbcoseven, suggests that the inclusion of Pr in general generates a susceptibility for a 1/3 super-lattice structure.

As \pbco\ is a stoichiometric compound, structure refinement is, in principle, possible from the large reciprocal scattering volume collected. We leave such a detailed analysis for a future communication. Instead, we bring forward symmetry arguments and the associated symmetry allowed distortion modes $u_i$.
Given that the ordering resonates at the Pr site, this atomic site is 
certainly involved in the super-lattice structure. The $Q=(\delta_1, 0, 0)$ ordering 
implies an in-plane distortion $u=(u_x,0,0,)$ of the Pr atoms. 
Given the average \textit{Pmmm} crystal structure of \pbco, the ordering vector allows four possible subspace groups - identified through the Isodistort software package~\cite{campbell_isodisplace_2006} (see supplementary table-II). From our diffraction data, it is not possible to differentiate bi-axial and twinned uniaxial orderings. Irrespectively of the order parameter symmetry, orthorhombic symmetry allowed space groups are \textit{Pmmm} and \textit{Pmm2}. For bi-axial symmetry two additional monoclinic subspace groups are allowed (\textit{P2/m} and \textit{Pm}). In all cases, in-plane Pr distortions modes are present. 
The causal relation between Pr-site distortions and mixed Pr-valence from oxygen hybridization remains to be clarified. It is, however, clear that the here observed $Q_1$-order will influence the hybridization within plane oxygen orbitals. As such, the reported super-structure links directly to the anomalous electronic properties of \pbco. We conclude that translational symmetry breaking at the Pr-site is detrimental to superconducting pairing.




\section*{Methods}
\textit{Film growth:}
Fully oxygenated, insulating \pbco\ (PBCO) films, 90 nm thick, were deposited on $5 \times 5$ mm$^2$ (001)-oriented SrTiO$_3$ substrates by radio-frequency (rf) sputtering. The deposition was carried out at a power of 50 W, a total pressure of 0.1 mbar (with an Ar:O$_2$ ratio of 4:1), and a substrate temperature of 800$^\circ$C. 
The \gypbco\ (Pr-YBCO) films, with a thickness of 120 nm, were grown on similar $5 \times 5$ mm$^2$ (001) SrTiO$_3$ substrates using pulsed laser deposition (PLD) at a heater temperature of 760$^\circ$C and an oxygen pressure of 0.6 mbar. After deposition, the films were slowly cooled to room temperature in an oxygen pressure of 700 mbar to ensure full oxygenation, resulting in Pr-YBCO with $y = 1$ \cite{arpaia2018probing}.
To obtain underdoped Pr-YBCO films with reduced chain oxygen content ($y = 0.67$), fully oxygenated samples were subjected to an $ex-situ$ annealing process \cite{arpaia2024engineering}. This consisted of 9 hours of treatment at 550$^\circ$C in a reduced oxygen atmosphere of 0.02 mbar.
The superconducting critical temperatures, $T_\mathrm{c}$, were determined from resistance versus temperature measurements, $R(T)$, performed using a PPMS DynaCool system (Quantum Design). The films exhibited $T_\mathrm{c}$ values of 53 K and 28 K for $y = 1$ and $y = 0.67$, respectively.
Structural characterization was carried out using X-ray diffraction. The out-of-plane reflections were used to extract the $c$-axis lattice parameter, which, combined with $T_\mathrm{c}$, was used to estimate the oxygen content $y$ \cite{arpaia2018probing,arpaia_signature_2023}. In-plane diffraction measurements confirmed the presence of the twinned orthorhombic \textit{Pmmm} crystal structure, consistent with the symmetry of the SrTiO$_3$ substrates, and indicative of the absence of long-range chain order in the films. 
Since the substrate has a tetragonal structure, all the investigated films present an almost perfect twinning.
The lattice constants of the investigated films are reported in the following table:
\begin{table}[ht!]
\begin{ruledtabular}
\begin{tabular}{ccccc}
 Compound & $d$ [nm] & $a$ [\AA] & $b$ [\AA]& $c$ [\AA] \\ 
\hline
  \pbco & 90 & 3.895 & 3.895 & 11.82 \\
  Y$_{0.7}$Pr$_{0.3}$Ba$_2$Cu$_3$O$_7$ &  60 & 3.858 & 3.907  &11.65\\
\ypbco & 120 & 3.877 & 3.897 & 11.73 \\
\end{tabular}
\end{ruledtabular}
\label{tab:tab1}	
\end{table}

\textit{Gracing-incidence x-ray diffraction:} 
High-energy gracing-incidence x-ray diffraction experiments were carried out at the P21.1 beam line~\cite{v_zimmermann_p211_2025} at the PETRA III synchrotron (DESY - Hamburg). The setup is essentially identical to that described in Ref.~\onlinecite{oppliger_discovery_2025}. 
The experimental results described here were obtained using a liquid nitrogen cryostat allowing for a 100~K base temperature.\\
The data were collected using an incident angle $\phi=0.030$° from the sample surface, chosen to maximize the signal from \gypbco\ films. The 3D dataset was then acquired by rotating the sample along the angle $\omega$ around the axis perpendicular to the beam axis in the range $30-230$°. The exposure time was $0.1$~s per frame.
The 1D profiles shown in Fig.~\ref{fig:figure1} were obtained by integrating the 3D datasets in $K$ and $L$ intervals of size $\pm0.05$ r.l.u.~and $\pm0.1$ r.l.u.~respectively.

\textit{Resonant X-ray Scattering (REXS):}
REXS measurements were performed at the UE46-PGM1 beam line~\cite{weschke_ue46_2018} at the BESSY II synchrotron. A helium cryostat permitted a base temperature of 10 K. $H$-$L$ momentum maps were acquired by rocking the incident angle $\theta$ at different scattering $\theta_{\text{sc}}$ angles. The $H,L$ values are then reconstructed for each point.
X-ray absorption spectroscopy (XAS) was performed in electron-yield mode and allowed us to identify the Pr-$M_5$ and Cu-$L_3$ resonances.
The $Q_1$ and $Q_2$ (CDW) peaks were extracted by fitting the spectra with a Gaussian plus a quadratic and linear backgrounds, respectively. All the spectra were normalized by the intensity of the incoming beam.


 


\section*{Acknowledgements}
We thank Caitlin Duffy and Cyril Proust for discussions of Pr-YBCO data. Our work has been supported by the Swiss National Science foundation on grant Nr.~200021\_188564. L.M acknowledges support from the Swiss National Science Foundation under Spark project CRSK-2\textunderscore220797.
L.M. and I.B. acknowledge support from University of Zurich under Postdoc Grants No. FK-23-128 and FK-23-113. 
J.O. acknowledges support from a Candoc grant of the University of Zurich (Grant no. FK-22-095).We acknowledge DESY (Hamburg, Germany), a member of the Helmholtz Association HGF, for the provision of experimental facilities.
Parts of this research were carried out at P21.1. Beamtime was allocated for proposal I-20240300 EC. This work was performed in part at Myfab Chalmers.\\

\section*{Authors contributions}
Quickly written. Can be missing contributions and contributions may come.
L.M. and S.R. contributed equally. The project was conceived by L.M. and thin films of \ypbco\ were grown by R.A. and F.L. Grazing incidence experiments were carried out by L.M., I.B., S.R., J.O., M.vZ., F.I., and J.C.. Resonant experiments were conducted by L.M., I.B., S.R., E.W., and J.C.. Data analysis was done by S.R., with input from L.M., J.O., and J.C.. The manuscript was written by L.M. and J.C. with inputs from all coauthors.
\bibliography{LQMRjohan, references}

\end{document}